\documentclass[12pt]{article}

\usepackage{amsmath,amsfonts}

\makeatletter

\@addtoreset{equation}{section}

\newtheorem{theorem}{Theorem}
\newtheorem{lemma}{Lemma}
\newtheorem{remark}{Remark}
\newtheorem{deff}{Definition}

\def\ov{\overline}
\def\A{{\bf A}}
\def\N{{\cal N}}
\def\P{\hbox{ Prob}}

\def\sumd{\displaystyle\sum}
\def\intd{\displaystyle\int}

\begin{document}

\title{Dynamical Behaviour of a Large Complex System}
\author{J. F. Feng
$^\dagger$ \qquad M. Shcherbina $^\ddagger$ \qquad B. Tirozzi $^*$
\\
$^\dagger$
Department of  Computer Science and Mathematics\\
 Warwick
University, Coventry
CV4  7AL, UK
\\
$^\ddagger$Institute for Low Temperature Physics, Ukr. Ac. Sci.,
47
Lenin Ave.\\
 Kharkov, Ukraine \\
$^*$Department of Physics, Rome University 'La Sapienza', 00185
Roma, Italy
 }
\date{}
\maketitle
\begin{abstract}
Limit theorems for a linear dynamical system with random
interactions are established. The theorems enable us to
characterize the dynamics of a large complex system in details and
assess whether a large complex system is weakly stable or unstable
(see Definition 1 below).

\end{abstract}

\section{Introduction}
Consider a dynamical system with random interactions (so-called a
complex system in \cite{[May]}) defined by
\begin{equation}
\ov x  '=-\kappa\ov x+{\A}\ov  x
 \label{dyn}\end{equation}
where $\ov x\in\mathbb{R}^n$, $\kappa$ is a real number and $\A$
is an $n\times n$ real random matrix with entries
\begin{equation} A_{ij}=n^{-1/2}W_{ij}.
  \label{A}\end{equation}
The question we ask here is {\it under what conditions on $\A$,
$\ov x$ is stable} as $n \to  \infty$.

Not surprisingly, this simple question has been extensively
discussed in the literature and has wide applications in various
areas. Early in 1972\cite{[May]}, Robert May 'answered' the
question in one of his Nature papers without proof. May's
arguments are based upon the asymptotical behaviour of the maximal
eigenvalue of the matrix $\A$. Using results related to Wigner's
semi-circle, he concluded that $\ov x$ is stable if
$$ \kappa > w$$
and unstable if
$$ \kappa <w$$
provided that $W_{ij},i,j=1,\cdots, n$ are i.i.d. random
variables, where $w$ is the finite standard deviation of $W_{ij}$.
In a nice paper published 12 years later  Cohen and Newman
\cite{[CN]}, after a careful investigation of various more complex
situations of the matrix $\A$, pointed out that May's criteria
above  could be false when $A$ does not vary with $n$ only by
scaling (not Eq. (\ref{A})). At the end of their paper (page 309),
they emphasized that the question asked at the beginning of the
current paper remains open. Furthermore  from May's criteria the
stability at the critical point $\kappa_c=w$ is not clear.

In the current paper, we aim to establish  limit theorems for $\ov
x$ and shed new lights into the issue discussed above.
 To facilitate
our discussion, we first introduce some notation. We are
interested in the statistical distribution of $\{x_i(t)\}_{i=1}^n$
on the real line.
 To study this distribution for any fixed time  $t$
 we define a normalized counting function of $x_i$.
\begin{equation}
\N_n(\lambda,t)=n^{-1}\sharp\{x_i(t)\le\lambda\}=n^{-1}\sum_{i=1}^n\theta(\lambda-x_i(t)),
 \label{N}\end{equation}
where $\theta(x)$ is a standard Heaviside function.  This function
is a  distribution of the random discrete measure
 on the real line. Our goal is to
study the behavior of this measure in the limit $n\to\infty$. More
precisely, we  prove that this measure becomes non-random, as
$n\to\infty$ (i.e. the variance of $\N_n(\lambda,t)$ tends to
zero) and the mean value coincides with the function
\begin{equation}
\lim_{n\to\infty}E\{\N_n(\lambda,t)\}=\int_{-\infty}^\lambda
dx\frac{e^{-(x-a(t))^2/2\sigma(t)}}{\sqrt{2\pi\sigma(t)}}.
 \label{erf}\end{equation}
This means that $\N_n(\lambda,t)$ becomes a normal distribution
with
 mean value $a(t)$ and  variance $\sigma(t)$. Hence we naturally introduce
 the following definition.
\begin{deff}
The dynamics $\ov x$ is (weakly) stable if and only if $\lim_{t\to
\infty} \sigma(t)<\infty$. \label{def}
\end{deff}

We present a necessary and sufficient condition for $\ov x$ to be
stable. When the matrix $\A$ has elements of i.i.d. random
variables in additional to some minor restrictions,
$\N_n(\lambda,t)$  converges to  the normal distribution with mean
$a(t)=\exp(-\kappa t)$ and variance
\begin{equation} \sigma(t)=e^{-2\kappa
t}\sum_{m=1}^\infty\frac{(wt)^{2m}}{m!m!}
 \label{sigma1}\end{equation}
where $\ov x(0)=(1,1,\cdots,1)$. Hence $\ov x(t)$ is stable if and
only if
$$\lim_{t\to \infty}\sigma(t)<\infty$$
The above results  are  then proved for the case of symmetric
matrix $\A$ ($\sigma(t)$ and $a(t)$ take   slightly more complex
forms) and generalized to arbitrary initial conditions.

Our proofs heavily rely on techniques recently developed in
mathematical physics, in particular in the treatment of the Spin
Glass model and the Hopfield model\cite{[hop]}. We first establish
that the system we consider here has a self-average property and
each single variable of the system is normally distributed. Based
upon these properties and the homogeneity of the system, our proof
is finally reduced to a simple calculation of  the mean and
variance of a single variable.

The applications of our theorems could be considerably wide, in
the current research interests of  network properties arising from
social science (actor networks, authors networks etc.), biology
(gene networks, protein networks, metabolic networks, and neuronal
networks etc.) and computer science (internet connections)
\cite{[pre]}. For example, we could directly apply our results to
networks of neurons, extend our results to networks where the
interactions have a long-tail distribution or are dependent such
as in small-world networks and scale-free networks. Locally, we
can extend our results to nonlinear dynamics which exhibit more
rich dynamical activities such as limit cycles and chaos
\cite{[91],[92]}.

\section{Results}
Let us consider the system of ordinary differential equations
defined by  Eq. (\ref{dyn}) and (\ref{A}) with $W_{ij}$ satisfying
conditions
\begin{equation}
E\{W_{ij}\}=0,\quad E\{W_{ij}^2\}=w^2
 \label{cond}\end{equation}
 and there exists $\alpha>0$ such that
\begin{equation}
\P\{|W_{ij}|>\lambda\}\le Ce^{-C\lambda^\alpha}, \quad (\forall
\lambda>0)
 \label{cond1}\end{equation}
 Supply  the system by the following initial conditions:
\begin{equation}
\ov x  (0)=\ov c,\quad  \ov c=(1,\dots,1)\in\mathbb{R}^n
 \label{in_c}\end{equation}
 Note that the solution $\ov x$ of the dynamics (\ref{dyn}) obviously
 depends on $n$. We drop the index $n$ whenever there is no
 confusion. We then have the following theorem.

 \begin{theorem}\label{thm:1}
 Consider the system (\ref{dyn}) with a matrix $\A$ of the form (\ref{A}) under conditions
 (\ref{cond})and (\ref{cond1}), and supply this system by the initial conditions (\ref{in_c}).
 Then for any $t>0$, $\N_n(\lambda,t)$ defined by (\ref{N}) converges in probability  to
  the normal distribution $\N(a(t),\sigma(t))$ (\ref{erf}) with the mean value
\begin{equation}
a(t)=e^{-\kappa t}
 \label{a}\end{equation}
 and  variance
\begin{equation}
\sigma(t)=e^{-2\kappa t}\sum_{m=1}^\infty\frac{(wt)^{2m}}{m!m!}.
 \label{sigma}\end{equation}
 \end{theorem}

\begin{remark} \label{rem:1_1}
We know that the series in the expression of $\sigma(t)$ is the
Bessel function and it behaves as $\exp(2wt)$. Therefore
$\kappa_c=w$, where $\kappa_c$ is the critical point of the
stability of the dynamics (\ref{dyn}). On the other hand, it is
readily seen from the expression of $\sigma(t)$ that when
$\kappa=w$ we have $\lim_{t\to \infty}\sigma(t)<\infty$. Hence
$\ov x$ is stable iff $\kappa_c\ge w$.
\end{remark}
\medskip

We study also the same problem in the real symmetric case, i.e.
the case when $\A$ is a real symmetric matrix ($A_{ij}=A_{ji}$ )
of the form (\ref{A}) and $W_{ij}$ ($i\le j$) are i.i.d. random
variables, satisfying conditions (\ref{cond}) and  (\ref{cond1}).

\begin{theorem}\label{thm:2}
 Consider the system (\ref{dyn}) with a real symmetric matrix $\A$ ($A_{ij}=A_{ji}$) of the form (\ref{A}) under conditions
 (\ref{cond})and (\ref{cond1}), and supply this system by the initial conditions (\ref{in_c}).
 Then for any $t>0$, $\N_n(\lambda,t)$ defined by (\ref{N}) converges in probability  to
  the normal distribution $\N(a_s(t),\sigma_s(t))$ (\ref{erf}) with the mean value
\begin{equation}
a_s(t)=\frac{1}{2\pi w}\int_{-2w}^{2w} \exp\{-\kappa t+\lambda
t\}\sqrt{4w^2-\lambda^2}d\lambda
 \label{a_s}\end{equation}
 and variance
\begin{equation}
\sigma_s(t)=\frac{1}{2\pi w}\int_{-2w}^{2w} \exp\{-2\kappa
t+2\lambda t\}\sqrt{4w^2-\lambda^2}d\lambda-\bigg(\frac{1}{2\pi
w}\int_{-2w}^{2w} \exp\{-\kappa  t+\lambda
t\}\sqrt{4w^2-\lambda^2}d\lambda\bigg)^2
 \label{sigma_s}\end{equation}
 \end{theorem}

Our last result is a generalizations of Theorems \ref{thm:1} and
\ref{thm:2} to the case of arbitrary initial distribution. More
precisely we study the system (\ref{dyn}) in both symmetric and
nonsymmetric cases with initial condition
\begin{equation}
\ov x(0)=\ov \xi =(\xi_1,\dots,\xi_n), \label{in_xi}\end{equation}
where $\{\xi_i\}_{i=1}^n$ are i.i.d. random variables independent
 of $\{W_{ij}\}_{i,j=1}^n$ with
\begin{equation}
E\{\xi_i\}=a_0,\quad E\{\xi_i^2\}=w_0^2\not=a_0^2, \quad E\{\xi_i^4\}\le C.
\label{cond_xi}\end{equation}

\begin{theorem}\label{thm:4}
 Consider the system (\ref{dyn}) with nonsymmetric matrix $\A$ ($A_{ij}\not=A_{ji}$) of the form (\ref{A}) under conditions
 (\ref{cond}) and (\ref{cond1}), and supply this system by the initial conditions (\ref{in_xi}) with (\ref{cond_xi}).
 Then for any $t>0$, $\N_n(\lambda,t)$ defined by (\ref{N}) converges in probability  to
  the distribution of the random variable of the form
\begin{equation}
y(t)=e^{-\kappa t}\xi_1+w_0\sigma^{1/2}(t) z
\label{final1}\end{equation}
where $z$ is a standard normal random variable independent of $\xi_1$ and $\sigma(t)$ is defined by (\ref{sigma}).

If $\A$ in (\ref{dyn}) is  a real symmetric matrix ($A_{ij}=A_{ji}$) of the form (\ref{A}) under conditions
 (\ref{cond}) and (\ref{cond1}), then under the initial conditions (\ref{in_xi}) and (\ref{cond_xi})
 and for any $t>0$, $\N_n(\lambda,t)$ (defined by (\ref{N})) converges in probability  to
  the distribution of the random variable of the form
\begin{equation}
y_s(t)=a_s(t)\xi_1+w_0\sigma^{1/2}_s(t) z,
\label{final2}\end{equation}
where $z$ is a standard normal random variable independent of $\xi_1$ and $a_s(t)$ and $\sigma_s(t)$ are defined by
(\ref{a_s}) and (\ref{sigma}) respectively.
 \end{theorem}

\begin{remark} \label{rem:1_2}
From results above, we see that in a sense our results are more
general than May's criteria. We actually completely characterize
the dynamical behaviour of $\ov x$, independent of whether it is
stable or not.
\end{remark}

\section{Proofs}

\begin{remark} \label{rem:1}
Let us observe  that the change of variables $\tilde
x_i(t)=e^{-\kappa t}x_i(t)$  allows us everywhere below consider
without loss of generality the system (\ref{dyn}) with $\kappa=0$.
\end{remark}
\medskip

\noindent {\it Proof of Theorem \ref{thm:1}}

The first  step is the proof of the self averaging property of $\N_n(\lambda,t)$, as
 $n\to\infty$, i.e. we prove that for any real $\lambda$
and $t>0$
\begin{equation}
\lim_{n\to\infty}E\bigg\{\bigg(\N_n(\lambda,t)-E\{\N_n(\lambda,t)\}\bigg)^2\bigg\}=0.
 \label{self-av_N}\end{equation}
According to the standard theory of measure, for this aim it is
enough to prove that  $g_n(z,t)$ -- the Stieltjes
transform of the distribution $\N_n(\lambda,t)$
\begin{equation}
g_n(z,t)=\int\frac{d\N_n(\lambda,t)}{\lambda-z}=n^{-1}\sum_{i=1}^n\frac{1}{x_i(t)-z},\quad(\Im
z\not=0 ),
 \label{g_n}\end{equation}
 for any $z:\Im z\not=0$ possesses  a self averaging
 property, i.e.
\begin{equation}
\lim_{n\to\infty}E\bigg\{\bigg|g_n(z,t)-E\{g_n(z,t)\}\bigg|^2\bigg\}=0
 \label{self-av}\end{equation}
 where $\Im
z$ is the imaginary part of $z$.  We prove (\ref{self-av})  by
using a standard method, based on the martingale differences. This
method was proposed initially in \cite{[FT],PS1} to prove the self
averaging property of the free energy of the
Sherrington-Kirkpatrick model of spin glasses. We use it in the
form:
\begin{theorem}\label{thm:3}
Consider the function $f(\ov\xi_1,\dots,\ov\xi_p)$
$f:\mathbb{R}^{\nu_1+\dots+\nu_p}\to\mathbb{C}$, where
$\ov\xi_1\in\mathbb{R}^{\nu_1},\dots,\ov\xi_p\in\mathbb{R}^{\nu_p}$
are independent random vectors. If for any $k=1,\dots,p$ there
exists a function
$\psi_k(\ov\xi_1,\dots,\ov\xi_{k-1},\ov\xi_{k+1},\dots\ov\xi_p)$
(independent of $\ov\xi_{k}$) and such that
\begin{equation}
E\bigg\{\bigg|f(\ov\xi_1,\dots,\ov\xi_p)-\psi_k(\ov\xi_1,\dots,\ov\xi_p)\}\bigg|^2\bigg\}\le
C_k,
 \label{t3.1}\end{equation}
 then
\begin{equation}
E\bigg\{\bigg|p^{-1}f(\ov\xi_1,\dots,\ov\xi_p)-E\{p^{-1}f(\ov\xi_1,\dots,\ov\xi_p)\}\bigg|^2\bigg\}\le
4p^{-2}\sum_{k=1}^p C_k.
 \label{t3.2}\end{equation}
\end{theorem}

\noindent {\it Proof of Theorem \ref{thm:3}}

 This theorem was proven in
\cite{P}, but since the proof is very simple we repeat it here for
the sake of completeness. Denote $E_k$ the averaging with respect
to the random vectors $\ov\xi_1,\dots,\ov\xi_{k}$, $E_p=E$ and
$E_0$ means the absence of averaging. Then it is evident that
\[
p^{-1}f(\ov\xi_1,\dots,\ov\xi_p)-E\{p^{-1}f(\ov\xi_1,\dots,\ov\xi_p)\}=
p^{-1}\sum_{k=1}^p\Delta_k,
\]
where
\[
\Delta_k=E_k\{f(\ov\xi_1,\dots,\ov\xi_p)\}-E_{k-1}\{f(\ov\xi_1,\dots,\ov\xi_p)\}.
\]
Since evidently for $k<j$ $E\{\Delta_k\ov\Delta_j\}=0$,
we obtain immediately that
\begin{multline*}
E\bigg\{\bigg|p^{-1}f(\ov\xi_1,\dots,\ov\xi_p)-E\{p^{-1}f(\ov\xi_1,\dots,\ov\xi_p)\}\bigg|^2\bigg\}
=
p^{-2}\sum_{k=1}^pE\{|\Delta_k|^2\}\\
\le
2p^{-2}\sum_{k=1}^pE\bigg\{\bigg|E_k\{f(\ov\xi_1,\dots,\ov\xi_p)-\psi_k(\ov\xi_1,\dots,\ov\xi_p)\}\bigg|^2\bigg\}\\+
2p^{-2}\sum_{k=1}^pE\bigg\{\bigg|E_{k-1}\{f(\ov\xi_1,\dots,\ov\xi_p)-\psi_k(\ov\xi_1,\dots,\ov\xi_p)\}\bigg|^2\bigg\}
\le 4p^{-2}\sum_{k=1}^p C_k.
\end{multline*}
Theorem \ref{thm:3} is proven. $\qquad \qquad \square$

Now we use Theorem \ref{thm:3} for the proof of (\ref{self-av}).
Then $p=n$, $\ov\xi_k=(W_{k1},\dots,W_{kn})$ and $f=ng_n(z,t)$.

Let us take
\[
\psi_k=\sum_{j=1}^n\frac{1}{x_j^{(k)}(t)-z},
\]
where $x_j^{(k)}(t)$ are the solutions of the system
\begin{equation}
\ov x  '=\A^{(k)}\ov  x,\quad \ov x(0)=\ov c,
 \label{dyn_k}\end{equation}
with the matrix $\A^{(k)}$, whose entries coincide with $A_{ij}$,
if $i\not=k$, $j\not=k$ and are equal to zeros otherwise. It is
evident, that $\psi_k$ does not depend on $(W_{k1},\dots,W_{kn})$.
So we are left to prove the bound (\ref{t3.1}). Due to the
symmetry of the problem it is enough to prove (\ref{t3.1}) for
$k=1$.

According to the standard theory of  differential equations,
considering the terms $A_{j1}x_1(t)$ as  known functions, we can
write for $j=2,\dots,n$
\begin{equation}
 x_j(t)=(e^{t\A^{(1)}}\ov
 c)_j+\int_0^t\sum_{i=2}^n(e^{(t-s)\A^{(1)}})_{ji}A_{i1}x_1(s)ds=
x_j^{(1)}(t)+\tilde\Delta_j(t).
 \label{repr}\end{equation}
 Let us  represent
 \[
\frac{1}{x_j^{(1)}(t)-z}-\frac{1}{x_j(t)-z}=\frac{\tilde\Delta_j(t)}{(x_j^{(1)}(t)-z)^2}-
 \frac{\tilde\Delta_j^2(t)}{(x_j^{(1)}(t)-z)^2(x_j(t)-z)}
 \]
 Then, using this representation for all terms of
 $(\psi^{(1)}-ng_n(z,t))$, except the first one, we write
 \begin{equation}
 \psi_1-ng_n(z,t)=\frac{1}{x_1^{(1)}(t)-z}-\frac{1}{x_1(t)-z}+I+II
 \label{t1.1}\end{equation}
 and
 \begin{multline}
E\{|I|^2\}=E\bigg\{\int_0^t\int_0^tds_1ds_2\sum_{j_1,j_2,i_1,i_2=2}^n\frac{(e^{(t-s_1)\A^{(1)}})_{j_1i_1}A_{i_11}}{(x_{j_1}^{(1)}(t)-z)^2}
\frac{(e^{(t-s_2)\A^{(1)}})_{j_2i_2}A_{i_21}}{(x_{j_2}^{(1)}(t)-z)^2}
x_1(s_1)x_1(s_2)\bigg\}\\
\le
C(t)E^{1/2}\bigg\{\int_0^tx_1^4(s)ds \,\bigg\}E^{1/2}\bigg\{\sum_{j_1,j_2,i_1,i_2=2}^n\sum_{j_1',j_2',i_1',i_2'=2}^n
\int_0^t\int_0^tds_1ds_2\\
\frac{(e^{(t-s_1)\A^{(1)}})_{j_1i_1}A_{i_11}}{(x_{j_1}^{(1)}(t)-z)^2}
\frac{(e^{(t-s_2)\A^{(1)}})_{j_2i_2}A_{i_21}}{(x_{j_2}^{(1)}(t)-z)^2}
\frac{(e^{(t-s_1)\A^{(1)}})_{j_1'i_1'}A_{i_1'1}}{(x_{j_1'}^{(1)}(t)-z)^2}
\frac{(e^{(t-s_2)\A^{(1)}})_{j_2'i_2'}A_{i_2'1}}{(x_{j_2'}^{(1)}(t)-z)^2}\bigg\}
 \label{t1.2}\end{multline}
 Here and  below we use notations $C(t)$ for some independent of $n$ positive functions, which
 satisfy the bound $C(t)\le e^{ct}$ with some positive $t$-independent constant $c$. These functions can be different
 in different formulas.

Now, since $\A^{(1)}$ and $x_{j}^{(1)}(t)$ do not depend on
$A_{i1}$,  the averaging with respect to all $A_{i1}$ gives us
that we have nonzero terms in the last sum only if
$i_1,i_2,i_1',i_2'$ are  pairwise equal, e.g., $i_1=i_1'$,
$i_2=i_2'$. Then, denoting
$$
D_{j}=\frac{1}{(x_j^{(1)}(t)-z)^2},
$$
after the summation with respect to $i_1,i_2,i_1',i_2'$ we get
\begin{multline}
E\{|I|^2\} \le
C(t)E^{1/2}\bigg\{\int_0^tx_1^4(s)ds \,\bigg\}E^{1/2}\bigg\{n^{-2}\sum_{j_1,j_2,j_1',j_2'=2}^n
\int_0^t\int_0^tds_1ds_2\\
(e^{(t-s_1)\A^{(1)T}}(t)e^{(t-s_1)\A^{(1)}})_{j_1'j_1}D_{j_1}D_{j_1'}
(e^{(t-s_2)\A^{(1)T}}e^{(t-s_2)\A^{(1)}})_{j_2'j_2}D_{j_2}D_{j_2'}\bigg\}\\
\le C(t)E^{1/2}\bigg\{\int_0^tx_1^4(s)ds
\,\bigg\}E^{1/2}\bigg\{n^{-2}| D |^4 e^{4t||\A^{(1)}||}\bigg\},
 \label{t1.3}\end{multline}
 where $\A^{T}$ means the transposed matrix of $\A$.
 Now  we use  the result of \cite{BdMS}, according to which under
condition  (\ref{cond1}) for Hermitian  matrix $\A$ with i.i.d.
complex elements, such that $A_{ij}=\overline A_{ji}$ and
$E\{A_{ij}\}=0$, $E\{|A_{ij}|^2\}=w^2$
\begin{equation}
 \P\{||\A||>2w+\varepsilon\}\le Ce^{-Cn^\gamma\varepsilon^{\gamma_1}},\quad \gamma=\frac{\alpha}{2(1+\alpha)},
 \quad\gamma_1=\frac{\alpha+6}{\alpha+4}
 \label{||A||}\end{equation}
 So, for non symmetric matrix $\A$ we can write $\A=\A_1+i\A_2$
 with $\A_1=\frac{1}{2}(\A+\A^*)$ and $\A_2=\frac{1}{2i}(\A-\A^*)$
 being Hermitian  matrices with i.i.d. elements, satisfying (\ref{cond1}).
 Then, since $||\A||\le ||\A_1||+||\A_2||$, we can derive from
 (\ref{||A||}) that in non symmetric case
\begin{equation}
 \P\{||\A||>4w+\lambda\}\le Ce^{-Cn^\gamma\lambda^{\gamma_1}},\quad \gamma=\frac{\alpha}{2(1+\alpha)},
 \quad\gamma_1=\frac{\alpha+6}{\alpha+4}
 \label{||A||_1}\end{equation}
This estimate is rather crude, because it is known that $||\A||\to
2w $, as $n\to\infty$ (see  \cite{GZ}, \cite{Sosh}, where the
large deviation type bounds was found for
$\P\{||\A\A^*||>4w^2+\varepsilon\}$ in the case $\alpha\ge 2$ or
\cite{ST1} for  the case $a_{ij}=w\pm1$). But it is enough for our purposes.

\medskip

\begin{remark}\label{rem:2}
 Inequality (\ref{||A||_1}) allows us to use $||\A||$ in our considerations like a
bounded random variables. Indeed, since, e.g., $|x_1(t)|\le
ne^{t||\A||}$, denoting
$P_n(\lambda)=\emph{\P}\{||\A||>4w+\lambda\}$ and using
(\ref{||A||_1}), we can write for any fixed $t$ and   $m,s<<n^\gamma/\log n$
\begin{multline*}
E\{|x_1(t)|^me^{s||\A||}\}\le
e^{s(4w+\epsilon)}E\{|x_1(t)|^m\theta(4w+\epsilon-||\A||)\}\\+
n^mE\{e^{(s+mt)||\A||}\theta(||\A||-4w-2\epsilon)\}\le
e^{s(4w+\epsilon)}E\{|x_1(t)|^m\}
+n^m\int_{\lambda>\epsilon}e^{(s+mt)\lambda}dP_n(\lambda)\\
\le
e^{s(4w+\epsilon)}E\{|x_1(t)|^m\}+O(e^{-Cn^{\gamma}\varepsilon/2})
\end{multline*}
Hence, below we use $||\A||$ as a bounded variable without additional explanations.
\end{remark}
\medskip

Using (\ref{||A||_1}) and the evident bound $|D_j|\le |\Im z|^{-2}$, we get
\begin{equation}
E\{|I|^2\}\le C(t)E^{1/2}\bigg\{\int_0^tx_1^4(s)ds \,\bigg\}.
 \label{t1.4}\end{equation}
 Besides, evidently
\begin{multline}
|II|\le |\Im z|^{-3}\sum_{j=2}^n\tilde\Delta_j(t)^2\\
=|\Im
z|^{-3}\int_0^t\int_0^tx_1(s_1)x_1(s_2)ds_1ds_2\sum_{i_1,i_2=2}^n(e^{(t-s_1)\A^{(1)T}}e^{(t-s_2)\A^{(1)}})_{i_1i_2}A_{i_11}A_{i_21}\\
\le|\Im z|^{-3}t\int_0^tx_1^2(s)ds \,
e^{2t||\A^{(1)}||}n^{-1}\sum_{i}^nW_{i1}^2.
\label{t1.4a}\end{multline}
Thus we get
\begin{equation}
E\{|II|^2\}\le C(t)E^{1/2}\bigg\{\int_0^tx_1^4(s)ds \,\bigg\}.
 \label{t1.5}\end{equation}
 Now we need the following lemma
\begin{lemma}\label{lem:1}
Under conditions of Theorem \ref{thm:1}
\begin{equation}
E\{x_1^4(t)\}\le C(t) \label{l1.1}\end{equation}
 and $x_1(t)$ converges in distribution to a Gaussian random variable.
\end{lemma}

\noindent {\it Proof of Lemma \ref{lem:1}}

Using the first equation in
(\ref{dyn_k}) for $k=1$ and the representation (\ref{repr}), we
get

\begin{equation}
x_1'(t)=\sum_{j=2}^n\frac{W_{1j}}{n^{1/2}}x_j^{(1)}(t)+\int_0^tds \,K_n(t-s)x_1(s),
 \label{t1.6}\end{equation}
where
\begin{equation}
K_n(t-s)=n^{-1}\sum_{i,j=2}^n(e^{(t-s)\A^{(1)}})_{ij}W_{1i}W_{j1}.
 \label{t1.7}\end{equation}
Hence
\begin{equation}
x_1(t)=\phi(t)+\int_0^tds \, S_n(t-s)x_1(s),
 \label{t1.8}\end{equation}
with
\[
\phi(t)=1+\sum_{j=2}^n\frac{W_{1j}}{n^{1/2}}d_j(t),\quad
d_j(t)=\int_0^tds \,x_j^{(1)}(s),\quad S_n(t)=\int_0^t d\tau\,
K_n(\tau)
\]
Making iteration in (\ref{t1.8}) we get
\begin{equation}
x_1(t)=\phi(t)+\sum_{m=1}^{n_1}\int_0^tds \,
S_n^{(m)}(t-s)\phi(s)+\int_0^tds \,
S_n^{(n_1+1)}(t-s)x_1(s),
 \label{t1.9}\end{equation}
where $n_1=[\log^2 n]$ ($[x]$ is the integer part of $x$) and $S_n^{(m)}(t)$ is defined as
\[
S_n^{(1)}(t)=S_n(t),\quad
S_n^{(m)}(t)=\int_0^tS_n(t-s)S_n^{(m-1)}(s) ds.
\]
Since evidently
\begin{equation}
|S_n(t)|\le
\frac{e^{t||\A^{(1)}||}-1}{||\A^{(1)}||}\bigg(n^{-1}\sum_{i=2}^nW_{1i}^2\bigg)^{1/2}\bigg(n^{-1}\sum_{i=2}^nW_{i1}^2\bigg)^{1/2}=\mathbf{K},
\label{t1.9a}\end{equation}
we have
\[
|S_n^{(m)}(t)|\le\frac{\mathbf{K}^m}{(m-1)!},
\]
and so for any $m\ge 2$
\[
\bigg|\int_0^tds \,
S_n^{(m)}(t-s)\phi(s)\bigg|\le\frac{\mathbf{K}^{m-1}}{(m-2)!}\bigg|\int_0^tds \,
(S_n(t-s))^2\bigg|^{1/2}\bigg|\int_0^tds \,\phi^2(s)\bigg|^{1/2}.
\]
Therefore
\begin{multline}
E\bigg\{\bigg(\int_0^tds \,
S_n^{(m)}(t-s)\phi(s)\bigg)^4\bigg\}\\
\le \frac{1}{((m-2)!)^4}E^{1/2}\bigg\{\bigg|\int_0^tds \,
(S_n(t-s))^2\bigg|^{4}\bigg|\int_0^tds \,\phi^2(s)\bigg|^{2}\bigg\}
 E^{1/2}\bigg\{\bigg|\int_0^tds \,\phi^2(s)\bigg|^{2}
\mathbf{K}^{4(m-1)}\bigg\}.
 \label{t1.10}\end{multline}
 But using definitions (\ref{t1.7}),(\ref{t1.8}) and taking into account that $\A^{(1)}$ and $\phi(t)$ do not depend on
 $A_{i1}$, we get for any $t$
 \begin{multline}
 E\bigg\{(S_n(t))^8\int_0^t\phi^4(s)ds\bigg\}=E\bigg\{\int_{0}^td\tau \,n^{-4}\bigg(\sum_{i,j=2}^n(e^{\tau\A^{(1)}})_{ij}
 W_{1i}W_{j1}\bigg)^8\int_0^t\phi^4(s)ds\bigg\}\\
 \le n^{-4}C(t)E\bigg\{\bigg(n^{-1}\sum_{i=2}W_{1i}^2\bigg)^4\int_0^t\phi^4(s)ds\bigg\}\le
 n^{-4}C(t)E\bigg\{\int_0^t\phi^4(s)ds\bigg\}.
 \label{t1.10a}\end{multline}
Hence, it follows from (\ref{t1.10}), (\ref{t1.10a}) and Remark \ref{rem:2} that for $m\le\log^2 n$
\begin{equation}
E\bigg\{\bigg(\int_0^tds \,
S_n^{(m)}(t-s)\phi(s)\bigg)^4\bigg\}\le n^{-2}
\frac{C^{4m}(t)}{((m-1)!)^4}E\bigg\{\int_0^t\phi^4(s)ds\bigg\}.
\label{t1.10b}\end{equation}
Similarly, using the trivial bound $|x_1(t)|\le ne^{t||\A||}$, we get
\begin{equation}
E\bigg\{\bigg(\int_0^tds \,
S_n^{(n_1+1)}(t-s)x_1(s)\bigg)^4\bigg\}\le n^{2}
\frac{C^{4(n_1+1)}(t)}{(n_1!)^4} \le O(n^{-2}).
\label{t1.10c}\end{equation}
Now, using (\ref{t1.10b}) and the H\"{o}lder inequality, we obtain
\begin{equation}
E\bigg\{\bigg(\sum_{m=1}^{n_1-1}\int_0^tds \,
S_n^{(m)}(t-s)\phi(s)\bigg)^4\bigg\}\le O(n^{-2})E\bigg\{\int_0^t\phi^4(s)ds\bigg\},
\label{t1.11a}\end{equation}
and so it follows from (\ref{t1.9}) and (\ref{t1.10c})
\begin{multline} E\{x_1^4(t)\}\le C E\{\phi^4(t)\}+O(n^{-2})\le C+
Cn^{-1}\sum_{j=2}^nE\bigg\{\bigg(\int_0^tds\, x^{(1)}_j(s)\bigg)^2\bigg\}+\\
C\bigg(n^{-1}\sum_{j=2}^nE\bigg\{\bigg(\int_0^tds \, x^{(1)}_j(s)\bigg)^2\bigg\}\bigg)^2+
Cn^{-2}\sum_{j=2}^nE\bigg\{\bigg(\int_0^tds \, x^{(1)}_j(s)\bigg)^4\bigg\}+O(n^{-2}).
\label{t1.11}\end{multline}
But
\begin{multline*}
n^{-1}\sum_{j=2}^nE\bigg\{\bigg(\int_0^tds \, x^{(1)}_j(s)\bigg)^2\bigg\}\le
tn^{-1}\int_0^tds \,\sum_{j=2}^nE\{(x^{(1)}_j(s))^2\}\\
\le
tE\bigg\{\int_0^tds \, n^{-1}\sum_{i,j=2}^n(e^{s\A^{(1)T}}e^{s\A^{(1)}})_{ij}\bigg\}
\le
tE\bigg\{\int_0^tds \,e^{2ts||\A^{(1)}||}\bigg\}\le C(t)
\end{multline*}
and (\ref{repr}) implies that
\[
E\{(x^{(1)}_j(t))^4\}\le
C(t)E\{x_j^4(t)\}+C(t)\int_0^tE\{x_1^4(s)\}ds \,
\]
Substituting these bounds in (\ref{t1.11}) and  taking into account that (due to the symmetry )
$E\{x_j^4(t)\}=E\{x_1^4(t)\}$, we get
\[
E\{x_1^4(t)\}\le C(t)+C(t)n^{-1}\int_0^tE\{x_1^4(s)\}ds
\]
So
\[
\max_{0\le s\le t}E\{x_1^4(s)\}\le C(t)+tC(t)n^{-1}\max_{0\le s\le t}E\{x_1^4(s)\}
\]
Hence, we have proved (\ref{l1.1}).

The second statement of Lemma \ref{lem:1} follows from representation (\ref{t1.9}), which now, using the bounds
(\ref{t1.10c}) and (\ref{t1.11a}), we rewrite as
\begin{equation}
x_1(t)=1+\sum_{j=2}^n\frac{W_{1j}}{n^{1/2}}d_j(t)+r_n(t),\quad
d_j(t)=\int_0^tds \,x_j^{(1)}(s),
\label{t1.12b}\end{equation}
where
\[
d_j(t)=\int_0^tds \,x_j^{(1)}(s),\quad E\{r_n^2(t)\}\le C(t)n^{-1}.
\]
Now we can apply the central limit theorem, because
 $d_j(t)$ are independent of $\{W_{1i}\}_{i=2}^n$ and, according to the above considerations,
\begin{multline}
n^{-1}\sum_{j=2}^nE\{(d_j(t))^4\}\le
C(t)n^{-1}\sum_{j=2}^n\int_0^tds \,E\{(x^{(1)}_j(s))^4)\}\\ \le
C(t)n^{-1}\sum_{j=2}^n\int_0^tds \, E\{x_j^4(s)\}=C(t)\int_0^tds
\,E\{x_1^4(s)\}\le C(t), \label{t1.12a}\end{multline} so $d_j(t)$
satisfy some kind of the Lindeberg condition. Lemma \ref{lem:1} is
proven.  $\qquad \qquad\square$

Using Lemma \ref{lem:1}, one can easily derive (\ref{t3.1})  from
(\ref{t1.1}), (\ref{t1.4}) and (\ref{t1.5}). Thus, we have proved
the self averaging of $g_n(z,t)$ (\ref{self-av}) and so also the
self averaging of $\N_n(\lambda,t)$ (\ref{self-av_N}).

Hence,
we need to study only $E\{\N_n(\lambda,t)\}$. But due to the
symmetry of the problem  it is easy to see that
\[
E\{\N_n(\lambda,t)\}=E\{\theta(\lambda-x_1(t))\}.
\]
So, $E\{\N_n(\lambda,t)\}$ coincides with the distribution $x_1(t)$. But, according to Lemma \ref{lem:1},
$x_1(t)$ converges in distribution, as $n\to\infty$, to a Gaussian random variable.
So, we are left only to find  the mean value
and the variance of $x_1(t)$.

 Using the bound (see the proof of Lemma \ref{lem:1})
\begin{multline}
E\bigg\{\bigg(\int_0^tds \,
S_n(t-s)x_1(s)\bigg)^2\bigg\}\\
\le E^{1/2}\bigg\{\int_0^tds \,
(S_n(t-s))^4\bigg\}
E^{1/2}\bigg\{\int_0^tx_1^4(s)ds \,\bigg\}\le C(t)n^{-1},
 \label{t1.13}\end{multline}
we derive from (\ref{t1.8}) that
\begin{equation}
E\{x_1(t)\}=1+O(n^{-1/2}),
 \label{t1.13a}\end{equation}
So we have proved (\ref{a}) for $\kappa=0$. Now using the remark in the beginning of the section, one can easily
get (\ref{a}) for any $\kappa\not=0$.

To prove (\ref{sigma}) define
\begin{equation}
R_n(t,s)=E\{x_1(t) x_1(s)\}=E\{x_j(t) x_j(s)\}.
\label{t1.12}\end{equation}
Using representation (\ref{t1.8}) for $x_1(t)$ and $x_1(s)$ and the bound (\ref{t1.13}),
we obtain
\begin{equation}
R_n(t,s)=1+\int_0^t\int_0^s dt'ds'R_n^{(1)}(t',s')+O(n^{-1}),
 \label{t1.14}\end{equation}
where we denote
\begin{equation}
R_n^{(1)}(t,s)=n^{-1}\sum_{j=2}^nE\{x_j^{(1)}(t) x_j^{(1)}(s)\}.
 \label{t1.15}\end{equation}
But from representation (\ref{repr}) and the inequality (\ref{t1.4a}) we get easily
\begin{multline*}
|R_n(t,s)-R_n^{(1)}(t,s)|
\le C(s)E^{1/2}\bigg\{n^{-1}\sum_{j=2}^n \tilde\Delta_j^2(t)\bigg\}
+C(t)E^{1/2}\bigg\{n^{-1}\sum_{j=2}^n \tilde\Delta_j^2(s)\bigg\}\\
\le (C(t)+C(s))n^{-1/2}.
\end{multline*}
Thus, we obtain from (\ref{t1.14}) the equation
\begin{equation}
R_n(t,s)=1+\int_0^t\int_0^s dt'ds'R_n(t',s')+O(n^{-1/2}),
 \label{t1.16}\end{equation}
 Iterating this equation, we find easily
\begin{equation}
\lim_{n\to\infty}R_n(t,s)=1+\sum_{m=1}^\infty
\frac{(wt)^m(ws)^m}{m!m!}.
 \label{t1.17}\end{equation}
 In particular,
\begin{equation}
\lim_{n\to\infty}E\{x_1^2(s)\}=1+\sum_{m=1}^\infty
\frac{(wt)^{2m}}{m!m!}.
 \label{t1.18}\end{equation}
 Now, using (\ref{t1.13a}), we get (\ref{sigma}) for $\kappa=0$. Then, using again the remark in the
 beginning of the section, it is easy to obtain (\ref{sigma}) for any $\kappa$.

Theorem \ref{thm:1}  is proven.   $\qquad \qquad\square$

\medskip

\noindent {\it Proof of Theorem \ref{thm:2}}

The first step here is again to prove the self averaging of $\N_n(\lambda)$, i.e. the proof of
(\ref{self-av_N}) or equivalently (\ref{self-av}). This proof almost coincides with that in Theorem \ref{thm:1}
and therefore we omit it. The only difference is in the proof of the analog of Lemma \ref{lem:1}.

\begin{lemma}\label{lem:2}
Under conditions of Theorem \ref{thm:1}
\begin{equation}
E\{x_1^4(t)\}\le C(t), \label{l2.1}\end{equation}
 and $x_1(t)$ converges in distribution to a Gaussian random variable.
\end{lemma}

\noindent {\it Proof of Lemma \ref{lem:2}}

As in the case of Lemma \ref{lem:1}, we use the equation, which can be obtained, if we
use the last $n-1$ equations to express $x_j(t)$ ($j=2,\dots,n$) via $x_1(t)$.
\begin{equation}
x_1'(t)=\sum_{j=2}^n\frac{W_{1j}}{n^{1/2}}x_j^{(1)}(t)+\int_0^tds \, K_n(t-s)x_1(s)ds,
 \label{l2.2}\end{equation}
where $x_j^{(1)}(t)$ are the solutions of (\ref{dyn_k}) in the symmetric case with $k=1$,
\begin{equation}\begin{array}{l}
K_n(t)=K_n^0(t)+\tilde K_n(t)\\
K_n^0(t)=n^{-1}\sumd_{i=2}^n(e^{t\A^{(1)}})_{ii}w^2,\\
\tilde K_n(t)=n^{-1}\sumd_{i,j=2,i\not=j}^n(e^{t\A^{(1)}})_{ij}W_{1i}W_{1j}+
n^{-1}\sumd_{i=2}^n(e^{t\A^{(1)}})_{ii}(W_{1i}^2-w^2),
\end {array}\label{l2.3}\end{equation}
and here and below  $A^{(1)}_{ij}$
coincides with $A_{ij}$, if $i\not=1,j\not=1$ and is equal to zero otherwise.
Hence
\begin{equation}
x_1(t)=\phi(t)+\int_0^tds \, S_n(t-s)x_1(s)
 \label{l2.4}\end{equation}
with
\begin{equation}\begin{array}{l}
\phi(t)=1+\sumd_{j=2}^n\frac{W_{1j}}{n^{1/2}}d_j(t),\quad
d_j(t)=\intd_0^tds \,x_j^{(1)}(s)ds,\\
 S_n(t)
=S_n^{0}(t)+\tilde S_n(t),\quad
 S_n^{0}(t)=\intd_0^t d\tau K_n^0(\tau),\quad
 \tilde S_n(t)=\intd_0^t d\tau \tilde K_n(\tau).
\end {array} \label{l2.4a}\end{equation}
Iterating (\ref{l2.4}) $n_1$ times ($n_1=[\log^2 n]$),
we get
\begin{equation}
x_1(t)=\phi(t)+\sum_{m=1}^{n_1}\int_0^t
S_n^{(m)}(t-s)\phi(s)ds+\int_0^t
S_n^{(n_1+1)}(t-s)x_1(s)ds,
 \label{l2.5}\end{equation}
where $S_n^{(m)}(t)$ is defined in (\ref{l2.9}) and has
 the same bound (\ref{t1.9a}).
Repeating the conclusions of Lemma \ref{lem:1}, we obtain finally
\begin{equation}\begin{array}{l}
x_1(t)=\phi(t)+\intd_0^tds \,
\hat S_n^0(t-s)\phi(s)ds+\int_0^t {\cal R}_n(t-s)\phi(s)ds+\tilde r_n(t),\\ \tilde r_n(t)=\intd_0^t
S_n^{(n_1+1)}(t-s)x_1(s)ds
 \end{array}\label{l2.6}\end{equation}
 where similarly to (\ref{t1.10c})
 \[  E\{\tilde r_n^4(t)\}\le O(n^{-2})
 \]
 and we denote
 \begin{equation}\begin{array}{l}
 \hat S_n^0(t)=
\sumd_{m=1}^{n_1} S_n^{(0,m)}(t),\\
S_n^{(0,1)}(t)=S_n^{0}(t),\quad S_n^{(0,m)}(t)=\intd_{0}^tS_n^{0}(t-s)S_n^{(0,m-1)}(s)ds
\end {array} \label{l2.6a}\end{equation}
and ${\cal R}_n(t)$ is the kernel of the remainder operator, which satisfies the bound
\begin{equation}
E\bigg\{\bigg(\int_0^tds \,
{\cal R}_n(t-s)\phi(s)\bigg)^4\bigg\}
\le  C(t) E^{1/2}\bigg\{\bigg|\int_0^tds \,
(\tilde S_n(t-s))^2\bigg|^{2}\bigg\}.
 \label{l2.7}\end{equation}
 Here and below we use the result of \cite{BdMS}, according to which in the symmetric case
 under conditions (\ref{cond}), (\ref{cond1}) the bound (\ref{||A||}) is valid.

 But, using definitions (\ref{l2.3}),(\ref{l2.5}) and taking into account that $\A^{(1)}$ does not depend on
 $A_{i1}$, we get for any $t$
 \begin{equation}
 E\left\{(\tilde S_n(t))^4\right\}\le C(t)n^{-2}.
\label{l2.8} \end{equation}
Hence, we derive from (\ref{l2.6}) and the fact that $\hat K_n^0(t)$ does not depend on $A_{i1}$ that
\begin{multline} E\{x_1^4(t)\}\le  C(t)
\bigg(n^{-1}\sum_{j=2}^nE\bigg\{\bigg(\int_0^tds \, x^{(1)}_j(s)\bigg)^2\bigg\}\bigg)^2\\+
C(t)n^{-2}\sum_{j=2}^nE\bigg\{\bigg(\int_0^tds \, x^{(1)}_j(s)\bigg)^4\bigg\}+O(n^{-2}).
\label{l2.9}\end{multline}
Then the bound (\ref{l2.1}) follows by the same way as in Lemma \ref{lem:1}.

\smallskip

The second statement of Lemma \ref{lem:2} follows from representation (\ref{l2.6}), by the same way
as in Lemma \ref{lem:1}, if we observe that
\begin{multline}
x_1(t)=\phi(t)+\int_0^tds \,
\hat S_n^0(t-s)\phi(s)ds+r_n(t)\\=1+\int_0^tds \,
\hat S_n^0(t-s)+\sum_{j=2}^n\frac{W_{1j}}{n^{1/2}}\bigg(d_j(t)+\int_0^tds \,
\hat S_n^0(t-s)d_j(s)\bigg)+r_n(t),
\label{l2.10}\end{multline}
where $d_j(t)$ and $\hat S_n^0(t)$ are independent of $\{W_{1i}\}{i=2}^n$, $\hat S_n^0(t)$ is bounded
and
\[E\{r_n^2(t)\}\le C(t)n^{-1}.\]
The analog of the Lindeberg condition follows from (\ref{t1.12a}).

Lemma \ref{lem:2} is proven. $\qquad \qquad\square$

Now,   the proof of the self averaging property of $g_n(z,t)$
(\ref{self-av}) and so also the self averaging property of
$\N_n(\lambda,t)$ (\ref{self-av_N}) is similar to the proof of
Theorem \ref{thm:1}.

Hence,
we need to study only $E\{\N_n(\lambda,t)\}$. But due to the symmetry of the problem,
 $E\{\N_n(\lambda,t)\}$ coincides with the distribution $x_1(t)$. And since, according to Lemma \ref{lem:2},
$x_1(t)$ converges in  distribution to a Gaussian random variable,
to prove Theorem \ref{thm:2} we are left to find
\begin{equation}\begin{array}{l}
a_{s,n}(t)=E\{(e^{t\A}\ov c)_1\}=E\{(e^{t\A})_{11}\}+\sumd_{j=2}^nE\{(e^{t\A})_{1j}\}\\
\sigma_{s,n}(t)=E\{x_1^2(t)\}-E^2\{x_1(t)\}
\end{array}\label{t2.3}\end{equation}
Let us use the Cauchy  formula, valid for any symmetric matrix $\A$,
\begin{equation}
(e^{t\A})_{1j}=\oint_L dz e^{zt}G_{1j}(z)dz,
\label{t2.4}\end{equation}
where $\mathbf{G}(z)=(\A-z)^{-1}$ is the resolvent of the matrix $A$ and the contour $L$
is taken in such a way to contain inside the interval $[-2w,2w]$, and the distance from $L$ to
$[-2w,2w]$ is more than some constant $d$. According to the result \cite{BdMS} (see (\ref{||A||})), then with probability
more than $1-e^{-Cdn^{\gamma}}$ all eigenvalues of $\A$ are inside the contour and the distance from any
of them to $L$ is more than $d/2$. Hence with the same probability formula (\ref{t2.4}) is valid,
and
\begin{equation}
\P\{||\mathbf{G}(z)||\le \frac{d}{2},\forall z\in L\}\ge 1-e^{-Cdn^{\gamma}}.
\label{normG}\end{equation}
We use also the following representation of the resolvent $\mathbf{G}(z)$:
\[
G_{1j}=\bigg(\sumd_{i,i'=2}^nG^{(1)}_{ii'}A_{1i}A_{1i'}+z\bigg)^{-1}\sumd_{i=2}^nG^{(1)}_{ji}A_{1i},
\quad (j\not=1),
\]
where $\mathbf{G}^{(1)}(z)=(\A^{(1)}-z)^{-1}$ is the resolvent of $\A^{(1)}$.
Hence, we can write
\begin{equation}
\sumd_{j=2}^nG_{1j}=\bigg(w^2\tilde g_n(z)+r_n(z)+z\bigg)^{-1}
\sumd_{i,j=2}^nG^{(1)}_{ji}A_{1i}, \label{t2.5}\end{equation}
where
\[\begin{array}{l}
\tilde g_n(z)=n^{-1}\sumd_{i=2}^nG^{(1)}_{ii},\\
r_n(z)=n^{-1}\sumd_{i=2}^nG^{(1)}_{ii}(W_{1i}^2-w^2)+n^{-1}\sumd_{i,i'=2,i\not=i'}^nG^{(1)}_{ii'}W_{1i}W_{1i'}.
\end{array}\]
Using that $\mathbf{G}^{(1)}(z)$ does not depend on $A_{1i}$, and (\ref{normG}) is valid also for
$||\mathbf{G}^{(1)}(z)||$, it is easy to get
\begin{multline}
E\{|r_n(z)|^2\}\le C n^{-2}E\bigg\{\sumd_{i=2}^n|G^{(1)}_{ii}(z)|^2\bigg\}
+
Cn^{-2}E\bigg\{\sumd_{i=2}^n(G^{(1)}(z)*G^{(1)}(\ov z))_{ii}\bigg\}\le C n^{-1}.
\label{t2.6}\end{multline}
Hence, it follows from (\ref{t2.5}), that
\begin{multline}
E\bigg\{\sumd_{j=2}^nG_{1j}\bigg\}=E\bigg\{(w^2\tilde
g_n(z)+z)^{-1}\sumd_{i,j=2}^nG^{(1)}_{ji}A_{1i}\bigg\}
\\
-E\bigg\{r_n(z)(w^2\tilde g_n(z)+z)^{-1}(w^2\tilde
g_n(z)+r_n(z)+z))^{-1}
\sumd_{i,j=2}^nG^{(1)}_{ji}A_{1i}\bigg\}=I-II
\label{t2.7}\end{multline} Since $\mathbf{G}^{(1)}(z)$ and
$g_n(z)$ do not depend on $A_{1i}$, $I=0$.  Besides, since
\[
|(w^2\tilde g_n(z)+z)^{-1}|,|(w^2\tilde g_n(z)+r_n(z)+z))^{-1}|\le ||\mathbf{G}||,
\]
 combining the Schwartz inequality with (\ref{t2.6}), we obtain
\[\begin{array}{l}
|II|\le C E^{1/2}\bigg\{n^{-1}\sumd_{i,j=2}^n\left(G^{(1)}(z)*G^{(1)}(\ov z)\right)_{ij}\bigg\}E^{1/2}\{|r_n(z)|^2\}
\le Cn^{-1/2}
\end{array}\]

So, it follows from (\ref{t2.3})-(\ref{t2.7}) that
\begin{equation}
\sum_{j=2}^nE\{(e^{t\A})_{1j}\}=O(n^{-1/2})
\label{t2.7a}\end{equation}
and so
\[
a_{s,n}(t)=E\{(e^{t\A})_{11}\}+O(n^{-1/2})=n^{-1}E\{\hbox{Tr }e^{t\A}\}+O(n^{-1/2}).
\]
Hence, according to the results of \cite{Wig}, we get
\[
\lim_{n\to\infty}a_{sn}(t)=a_{s}(t)=\frac{1}{2\pi w}\int_{-2w}^{2w} e^{\lambda
t}\sqrt{4w^2-\lambda^2}d\lambda
\]
and so we have proved (\ref{a_s}) for $\kappa=0$. Using remark in the beginning of the section,
now it is easy to obtain (\ref{a_s}) for $\kappa\not=0$.

To find $\sigma_{sn}(t)$ let us observe that, due to the symmetry,
\begin{multline*}
E\{x_1^2(t)\}=n^{-1}\sum_{i=1}^nE\{x_i^2(t)\}=n^{-1}\sum_{i=1}^nE\{(e^{tA}\ov c)_i^2\}\\
=
n^{-1}\sum_{i,j=1}^nE\{(e^{2tA})_{ij}\}=E\{(e^{2tA}\ov c)_1\}=E\{x_1(2t)\}
\end{multline*}
Now it is easy to obtain (\ref{sigma_s}) for any $\kappa$.

\medskip

\noindent {\it Proof of Theorem \ref{thm:4}}

The proof of the fact that  $\N_n(\lambda,t)$ is a self averaging quantity and coincides in the limit in the
distribution of $x_1(t)$ is the same as in Theorem \ref{thm:1}, \ref{thm:2}. Thus we are left to prove only that
$x_1(t)$ can be represented in the form (\ref{final1}) in the non symmetric case or (\ref{final2}) in the symmetric case.

In the non symmetric case we get similarly to (\ref{t1.12b}), that
\begin{equation}
x_1(t)=\xi_1+\sum_{j=2}^n\frac{W_{1j}}{n^{1/2}}d_j(t)+r_n(t),\quad
d_j(t)=\int_0^tds \,x_j^{(1)}(s),
\label{t4.1}\end{equation}
where
\[
d_j(t)=\int_0^tds \,x_j^{(1)}(s),\quad E\{r_n^2(t)\}\le C(t)n^{-1}
\]
and  since $d_j(t)$ are independent on $W_{1j}$ and $\xi_1$ and
satisfy the inequality(\ref{t1.12a}, we obtain that the second sum
converges in probability to a normal random variable with  zero
mean and the variance
\begin{equation}
\sigma^\xi(t)=\lim_{n\to\infty}n^{-1}\sum E\{d_j^2(t)\}
\label{t4.2}\end{equation}
Now, let us denote
\[
R_n(t,s)=E\{x_1(t)x_1(s)\}.
\]
 Then repeating  the conclusions (\ref{t1.14})-(\ref{t1.18}) of Theorem \ref{thm:1}, we get from (\ref{t3.1})
\begin{equation}
R(t,s)=\lim_{n\to\infty}R_n(t,s)=w_0^2\left(1+\sum_{m=1}^\infty \frac{t^ms^m}{m!m!}\right)
\label{t4.3}\end{equation}
Hence,  by (\ref{t4.2}) and the symmetry of the problem, we get
\begin{equation}
\sigma^\xi(t)=\int_0^t \int_0^tdt_1dt_2R(t_1,t_2)=w_0^2\sum_{m=1}^\infty \frac{t^{2m}}{m!m!}
\label{t4.4}\end{equation}
So, we have proved (\ref{final1}) in the case $\kappa=0$. Then, using Remark \ref{rem:1}, we obtain (\ref{final1})
for any $\kappa$.

\medskip

To prove (\ref{final2}) we use the analog of (\ref{l2.10}) which in the case of (\ref{in_xi}) has the form
\begin{multline}
x_1(t)=\xi_1\left(1+\int_0^tds \,
\hat S_n^0(t-s)\right)+\sum_{j=2}^n\frac{W_{1j}}{n^{1/2}}\bigg(d_j(t)+\int_0^tds \,
\hat S_n^0(t-s)d_j(s)\bigg)+r_n(t)\\=s_n(t)\xi_1+z_n+r_n(t),\quad E\{r_n^2(t)\}\le C(t)n^{-1},
\label{t4.5}\end{multline}
where  $d_j(t)$ and  $\hat S_n^0(t)$   are independent of $\{W_{1i}\}_{j=2}^n$, $\hat S_n^0(t)$ is bounded and $d_j(t)$
satisfy (\ref{t1.12a}). Thus, according to the central limit theorem,
$z_n$ converges in distribution to a Gaussian random variable, independent of $\xi_1$.
Besides, $s_n(t)$
 is a self averaging quantity. To prove this it is enough to prove that $\hat S_n^0(t)$ is a self averaging quantity.
 The last statement follows from  the representation (\ref{l2.6a}), if we know that
 $ S_n^0(t)$ is a self averaging  quantity.   But,
 by definitions (\ref{l2.4a}) and (\ref{l2.3}) and  the spectral theorem,
 \[S_n^0(t)=\int_0^td\tau n^{-1}\hbox{Tr}e^{\tau \A^{(1)}}=\int_0^td\tau \int e^{\lambda\tau}d\N_n^*(\lambda)
 \]
where
\[\N_n^*(\lambda)=n^{-1}\sum_{i=1}^n\theta(\lambda-\lambda_i^*)
\]
is a normalized counting measure of  eigenvalues of $\A^{(1)}$.
So,  the self averaging  of $s_n(t)$ follows from the self
averaging of $\N_n^*(\lambda)$, which is a well known result (see,
e.g. \cite{Wig} or the review paper \cite{P}).

Thus, to finish the proof of (\ref{final2}) we are left to find  $E\{s_n(t)\}$ and the variance of $z_n$
in (\ref{t4.5}). But,  (\ref{t4.5}) implies that
\begin{equation}
E\{s_n(t)\}=w_0^{-2}E\{x_1(t)\xi_1\}+O(n^{-1/2}), \quad E\{z_n^2\}=E\{x_1^2(t)\}-E^2\{s_n(t)\}E\{\xi_1^2\}+O(n^{-1})
\label{t4.6}\end{equation}
So, using the fact that $x_1(t)$ is a solution of (\ref{dyn}) with the initial condition (\ref{in_xi}), we get
\begin{multline}
E\{x_1(t)\xi_1\}=E\{\xi_1^2\}E\{(e^{t\A})_{11}\}+\sum_{j=2}^nE\{(e^{t\A})_{1j}\}E\{\xi_1\}E\{\xi_j\}\\
=E\{\xi_1^2\}E\{(e^{t\A})_{11}\}+E^2\{\xi_1\}\sum_{j=2}^nE\{(e^{t\A})_{1j}\}
\label{t4.7}\end{multline}
No, using (\ref{t2.7a}), we get
\[
E\{s_{n}(t)\}=E\{(e^{t\A})_{11}\}+O(n^{-1/2})=n^{-1}E\{\hbox{Tr }e^{t\A}\}+O(n^{-1/2}).
\]
Hence, according to the results of \cite{Wig}, we get
\[
\lim_{n\to\infty}E\{s_{n}(t)\}=\frac{1}{2\pi w}\int_{-2w}^{2w}
e^{\lambda t}\sqrt{4w^2-\lambda^2}d\lambda.
\]
To compute $E\{z_n^2\}$, we write, using  that $x_1(t)$ is a solution of (\ref{dyn}) with  (\ref{in_xi})
and taking into account the symmetry of the problem,
\begin{multline}
E\{x_1(t)^2\}=\sum_{i=1}^nE\{\xi_i^2\}E\{(e^{t\A})_{1i}^2\}+\sum_{i,j=1,i\not=j}^nE\{(e^{t\A})_{1i}(e^{t\A})_{1j}\}E\{\xi_i\}E\{\xi_j\}\\
=E\{\xi_1^2\}E\{(e^{2t\A})_{11}\}+E^2\{\xi_1\}n^{-1}\sum_{i,j=1,i\not=j}^n\sum_{k=1}^nE\{(e^{t\A})_{ik}(e^{t\A})_{kj}\}\\
=w_0^2E\{(e^{2t\A})_{11}\}+E^2\{\xi_1\}n^{-1}\sum_{i,j=1,i\not=j}^nE\{(e^{2t\A})_{ij}\}.
\label{t4.8}\end{multline}
But, according to (\ref{t2.7a}) the second sum in the r.h.s of(\ref{t4.8}) is $O(n^{-1/2})$. And so, using the above consideration,
we have
\[
\lim_{n\to\infty}E\{x_1^2(t)\}=E\{\xi_1^2\}\lim_{n\to\infty}n^{-1}E\{\hbox{Tr
}e^{2t\A}\}=\frac{w_0^2}{2\pi w}\int_{-2w}^{2w} e^{2\lambda
t}\sqrt{4w^2-\lambda^2}d\lambda.
\]
Finally, we obtain
\begin{equation}
\lim_{n\to\infty}E\{z_n^2\}=\frac{w_0^2}{2\pi w}\int_{-2w}^{2w}
e^{2\lambda
t}\sqrt{4w^2-\lambda^2}d\lambda-w_0^2\left(\frac{1}{2\pi
w}\int_{-2w}^{2w} e^{\lambda
t}\sqrt{4w^2-\lambda^2}d\lambda\right)^2=w_0^2\sigma_{s}(t).
\label{t4.9}\end{equation} Now, relations
(\ref{t4.5})-(\ref{t4.9}) imply (\ref{final2}) for $\kappa=0$.
Then, using Remark \ref{rem:1}, we  obtain (\ref{final2}) for any
$\kappa$.

\noindent {{\bf Acknowledgment.}  We are grateful to one of the
Associate Editors for his/her discussion on the manuscript. J.F.
was partially supported by grants from UK EPSRC(GR/R54569),
(GR/S20574),  and (GR/S30443).}

\end{document}